\documentclass[12pt]{article}
\input epsf.sty
\def\be{\begin{equation}} \def\ee{\end{equation}}
\def\bea{\begin{eqnarray}} \def\eea{\end{eqnarray}} \def\ba{\begin{array}}
\def\ea{\end{array}} \def\ben{\begin{enumerate}} \def\een{\end{enumerate}}
  
\def\lll{\label}
\newcommand{\eqn}[1]{(\ref{#1})}
\newcommand{\npb}[3]{Nucl. Phys. {\bf B#1} ({#2}) {#3}}

\newcommand{\hepth}[1]{{\tt hep-th/{#1}}}

\def\br{\nonumber\\}

\begin{document}
{}~
\hfill\vbox{\hbox{hep-th/yymmnnn} \hbox{SINP/TNP/07-01}}\break

\vskip .5cm
\centerline{\large \bf
3-Branes on Eguchi-Hanson $6D$ Instantons}
\vskip .5cm

\vspace*{.5cm}

\centerline{  Harvendra Singh}

\vspace*{.25cm}

\centerline{ \it Saha Institute of Nuclear Physics} 
\centerline{ \it  1/AF Bidhannagar, Kolkata 700064, India}

\vspace*{.5cm}

\vskip.5cm
\centerline{E-mail: h.singh[AT]saha.ac.in }

\vskip1cm
\centerline{\bf Abstract} \bigskip

We use the approach used by Eguchi-Hanson in constructing four-dimensional
instanton metrics
and construct a class of regular 
six-dimensional instantons which are nothing but $S^2\times S^2$ 
resolved conifolds. 
 We then also obtain D3-brane solutions on these EH-resolved conifolds.

\vfill \eject

\baselineskip=16.2pt

\section{Introduction}

 The AdS/CFT conjecture 
\cite{maldacena,gubser,witten}
  relates string theory on anti-de Sitter bulk spacetime 
to a four-dimensional conformal 
gauge theory on the boundary. It serves as being one of the concrete 
realisation of the holographic idea \cite{holo}. In this picture the 
$SU(N)$ 
super-Yang-Mills theory lives on the 
boundary  of the bulk $AdS_5\times S^5$ space. Since then
this gauge-gravity duality has been extensively tested and has 
provided a fruitful alternative in understanding 
of various aspects of gauge theories, both qualitatively and 
quantitatively. The large  body of work is cited in some 
reviews, for example \cite{aharony}.  

It has been rather interesting to study AdS/CFT for the class of manifolds
$AdS_5\times T^{1,1}$ which have less supersymmetry, where $T^{1,1}$ is an 
Einstein space. This requires 
constructing D3-brane solutions over $M_4\times Y^{p,q}$ spaces where   
$Y^{p,q}$ is the  six-dimensional  Calabi-Yau cone. These cone-like 
geometries 
are singular at the tip of the cone, but  
deformations on these conifold geometries can be performed \cite{ossa}
so that they become regular Calabi-Yau geometries. 
Several new resolved solutions have recently appeared in literature 
\cite{klebanov,zayas}. 

In this short note we first construct a regular Calabi-Yau cone solution
which we obtain by adopting standard method of
constructing Eguchi-Hanson 
$4D$ instantons. We call them EH-resolved cone as they share some unique
 properties with  Eguchi-Hanson instantons. But there 
are standard
techniques to resolve these cones, see \cite{ossa,chen1}. 
We then look for D3-brane solutions on 
these EH-resolved cones. The paper is organised as follows. 
In the section-2, we 
review Eguchi-Hanson $4D$ instanton geometry. In section-3 we construct 
six-dimensional regular cone-like geometry using Eguchi-Hanson ansatz. 
We then obtain D3-brane 
configurations over these spaces in the section-4. 
The results are summarised in the 
last section.  
 
\section{The  $4D$ EH instantons}
The Euclidean gravity solutions with finite action and a self-dual 
 curvature are manifolds that are generally classified as gravitational 
instantons.  The gravitational fields are 
localised in space and the metric becomes asymptotically locally flat at 
infinity. \footnote{It has been recently found that the 
information-geometry on the solution space of these instantons 
has a constant negative curvature \cite{ps}. 
The results are found to be quite  
similar to the information-metric case of Yang-Mills 
instantons  \cite{blau}.} 
In this section we review the main 
aspects of the Eguchi-Hanson $4D$ solutions. 

In the case of the Eguchi-Hanson (EH) manifold the metric ansatz is 
\cite{eguchi} 
\be\label{eh}
ds_{EH}^2={dr^2\over g(r)}  +{r^2\over 4}
(d\theta^2+ \sin^2\theta d\phi^2)+{r^2\over 4}g(r)
(d\psi+ \cos\theta d\phi)^2
\ee
where $r^2=t^2+x^2+y^2+z^2 $. The $t,x,y,z$ are Cartesian 
coordinates
which define the base space  which is $R^4$. There is 
a complete radial 
symmetry and one is free to pick the center in the base space. 
The above metric solves Euclidean gravity equations in flat space provided 
$g(r)$ satisfies a monopole like first order equation
\be
g'(r)+{4\over r}(g -1)=0 
\ee
which has an immediate solution \cite{eguchi}
\be
g(r)=1-{a^4\over r^4}, ~~~~~(r\ge a)
\ee
where $a$ is an integration constant.
  By redefining the coordinate $\rho^4=r^4-a^4$
one can also express metric as
\cite{eguchi}
\be\label{eh2}
ds_{EH}^2=( 1+{a^4\over \rho^4})^{-{1\over2}}(d\rho^2 +(\rho^2/4)
(d\psi+\cos\theta d\phi)^2)+ 
(1+{a^4\over \rho^4})^{1\over2} {\rho^2\over4}(d\theta^2+ \sin^2\theta 
d\phi^2)
\ee
with the  coordinate ranges 
\be\label{ranges1}
0\le \rho\le\infty,~0\le\theta\le\pi,~0\le\phi\le 2\pi,~0\le\psi\le 2\pi .
\ee
A non-singular Maxwell field with self-dual field strength can be 
introduced in the Eguchi-Hanson background
\bea
&&A_{(1)}= {a^2\over \sqrt{\rho^4+a^4}}\sigma_z \, ,\br
&&
F_{(2)}={2 a^2\over \rho^4+a^4}(e^3\wedge e^0 + e^1\wedge e^2).
\eea
Thus for EH instanton,  
\be
(F_{\mu\nu})^2={16 a^4\over (\rho^4+a^4)^2} \ .\label{u2}
\ee
Though $a$ is an integration constant, but it is related to 
the presence of Maxwell field here.
For the instantons located away from the origin in $R^4$,
one can
introduce 4 new parameters, $x_0^i$, as the position variables. This is
 possible because the EH metric and the gauge background have complete
radial symmetry, so  $\rho^2=|x-x_0|^2$. However, in any case the center 
coordinates, $x_0^i$, of the EH metric 
\eqn{eh2} and the gauge field \eqn{u2} must coincide in order that the 
solution exists. It is interesting to note that the  Eq.\eqn{u2} is 
strikingly similar to the Yang-Mills instanton
 field strength in \cite{instantons,dorey,blau}, 
the only difference being 
the distribution of powers in the denominator. This fact did lead us to 
conclude 
that the information metric approach \cite{blau} to holography 
can also be studied for 
this class of curved EH instantons also; see
for more details \cite{ps}.

\section{ A EH-resolved $6D$ conifold}
We construct a  class of regular six-manifolds
which  are the Ricci-flat  K\"ahler solutions of the  Einstein 
equations. 
For this purpose, we  know the existence of 
the conifold geometries \cite{ossa,klebanov-witten} and 
corresponding $AdS/CFT$ analysis has been studied in detail by  
\cite{klebanov,zayas}. The many deformed and resolved conifolds 
have also been worked out in \cite{zayas,chen1} previously.  
As we are interested in Eguchi-Hanson class of instantons on the conifold,
so we look for a metric ansatz
resembling the Eguchi-Hanson $4D$ metric 
\be\label{bh1}
ds_{6}^2={dr^2\over f(r)} 
+{r^2\over 6}\left( 
(d\theta_1^2+ \sin^2\theta_1 d\phi_1^2)+
(d\theta_2^2+ \sin^2\theta_2 
d\phi_2^2) \right)+ {r^2\over 9}f(r) 
(d\psi+ \sum_{i=1}^2\cos\theta_i d\phi_i)^2 \ .
\ee
The angular coordinate ranges are taken as
\be\label{bh2}
~0\le\theta_i\le\pi,~0\le\phi_i\le 2\pi,~0\le\psi\le 2\pi .
\ee
One  will easily note that the  
$r=constant$ sections of the above metric are indeed $T^{1,1}$ geometry, 
which is an Einstein space with  positive curvature. 
Now all the $R_{\mu\nu}=0$ equations are solved by the 
metric \eqn{bh1} provided
$f(r)$ satisfies the  following monopole type  equation,
\be\label{bh3}
f'+{6\over r}(f-1)=0. \label{mon1}
\ee
Note that this first order equation appears in various instanton
solutions including the Yang-Mills and the $4D$ Eguchi-Hanson one. 
These are consequences of BPS conditions.
The above differential equation has one trivial solution  
$f=1$, in which case the 
metric becomes that of the  singular Calabi-Yau cone over the base  $T^{1,1}$ 
\cite{ossa}
\be\label{bh4}
ds_6^2=dr^2 + r^2 ds_{T^{1,1}}^2
\ee 
which is Ricci-flat and the  K\"ahler 2-form is given by 
\be\label{bh5}
J={r^2\over6}(\sin\theta_1d\theta_1\wedge 
d\phi_1+\sin\theta_2d\theta_2\wedge 
d\phi_2) +{r\over3}(d\psi+ 
\cos\theta_i d\phi_i)\wedge dr\ .
\ee
 The nontrivial solution of \eqn{bh3}  is
\be \label{fgh}
f(r)=1-{a^6\over r^6}\ ,~~~~~~a\le r <\infty 
\ee
where $a$ is an integration constant. 
We can take it to be the 
size of the instantons, or the measure of deformation. \footnote{ The third 
solution with plus sign 
$f(r)=1+{a^6\over r^6}$ will have a naked singularity at $r=0$.}
For the solution \eqn{fgh} the metric becomes 
\be\label{bh6}
ds_{6}^2={dr^2\over 1-{a^6\over r^6}} 
+{r^2\over 6}\left( 
(d\theta_1^2+ \sin^2\theta_1 d\phi_1^2)+
(d\theta_2^2+ \sin^2\theta_2 
d\phi_2^2)\right)+ {r^2\over 9}(1-{a^6\over r^6}) (d\psi+ \cos\theta_i 
d\phi_i)^2 .
\ee
which appears quite the same as Eguchi-Hanson metric \eqn{eh}.
This EH-resolved six-manifold is topologically $R^{+}\times S^2\times 
S^3$. An interesting {\it unique} property of the EH-resolved metric, that
is why we call it so, is that the determinant of the resolved metric remains
the same as the
unresolved metric. It was also the case with the $4D$ instantons. 
The apparent coordinate singularity at $r=a$ is 
removable, it can be seen as follows. Near $r=a$ we 
define new radial coordinate $u^2 = r^2(1-{a^6\over r^6})$, so the metric 
\eqn{bh6} can be written as
\be\label{bh6a}
ds_{6}^2={du^2 \over(1+{2a^6\over r^6})^2} +{u^2\over9}(d\psi+ 
\cos\theta_i 
d\phi_i)^2
+{r^2\over 6}\left( 
(d\theta_1^2+ \sin^2\theta_1 d\phi_1^2)+
(d\theta_2^2+ \sin^2\theta_2 
d\phi_2^2)\right) .
\ee
So near $r=a$ it becomes
\be\label{bh6b}
ds_{6}^2\approx{1\over 9} (du^2  + u^2(d\psi+ 
\cos\theta_i 
d\phi_i)^2)
+{a^2\over 6}\left( 
(d\theta_1^2+ \sin^2\theta_1 d\phi_1^2)+
(d\theta_2^2+ \sin^2\theta_2 
d\phi_2^2)\right) \ ,
\ee
and thus the metric can be 
made regular with the topology 
being $R^2\times S^2\times S^2$ near $r=a$, if the range of  
$\psi$ is restricted to $0\le\psi\le 2\pi$. While for  $r\gg a$ the 
metric \eqn{bh6} 
becomes the metric over the K\"ahler cone \eqn{bh4}. 

In any case the K\"ahler 
2-form for the metric \eqn{bh6} is  \footnote{ Note that it can 
be written as $J \equiv d A = d( -{r^2\over6} (d\psi+ \cos\theta_i 
d\phi_i))$. Also $J\wedge J\wedge J\equiv Vol[M_6] $.}
\be
J=e^5\wedge e^0+e^1\wedge e^2+ e^3\wedge e^4
\ee
The vielbeins are
\bea
&&e^0=( 1-{a^6\over r^6})^{-1/2}dr,~e^1= {r\over\sqrt{6}} d\theta_1,~ 
e^2={r\over\sqrt{6}} \sin\theta_1d\phi_1, e^3= {r\over\sqrt{6}} 
d\theta_2,~ \br
&& e^4={r\over\sqrt{6}} \sin\theta_2d\phi_2 , 
~e^5={r\over3} (1-{a^6\over 
r^6})^{1/2} (d\psi+ \cos\theta_1 d\phi_1 +\cos\theta_2 d\phi_2) 
\eea
 
\subsection{Adding the tensor fields}
We now wish to introduce a pair of second rank tensor fields $B_{\mu\nu}$
and $C_{\mu\nu}$
in the 6-dimensional manifold \eqn{bh6}. The most plausible combined 
action is
\be
\int \left( R \ast 1 -{1\over 2!} H_{(3)}\wedge\ast H_{(3)}
-{1\over 2!} F_{(3)}\wedge\ast F_{(3)} 
-{1\over 2!} F_{(1)}\wedge\ast F_{(1)} 
+ {\rm boundary-terms} \right)
\ee
where field strengths are given by $H_{(3)}=d B_{2}$ and $F_{(3)}=d 
C_{2}-\chi dB_2$ and $F_{(1)}=d\chi$. In our Hodge-dual 
convention: $\ast 
1=e^1\wedge \cdots\wedge e^6 \equiv \sqrt{g} [d^6x].$ 
The field 
equations which follow from the above action are
\bea
 &&d \ast F_3 =0\ , ~~~~ d (\ast H_3 -\chi \ast F_3)=0\ . \br
&&d\ast d \chi= dB_2\wedge\ast dC_2 
\eea
Thus a  vanishing $\chi$ solution exists provided we ensure 
$F_{\mu\nu\lambda} H^{\mu\nu\lambda}=0$. For $\chi=0$
the tensor field equations are readily solved provided we take 
\be\label{dual}
\ast F_3 =- H_3\ , ~~~~ \ast H_3 = F_3 ~.
\ee
From here, one can construct a complex harmonic $(2,1)$-form 
$G_3=F_3+i H_3$ which satisfies the following complex self-duality 
relation 
\be\label{selfdual}
\ast G_3= i G_3.
\ee
and its energy-momentum tensor identically vanishes.

\subsection{The singular $B$-field  solution}

We now construct a solution    
of the coupled field equations of Einstein and 2-rank tensor 
fields. It is found that a solution exists 
where the metric is taken to be the Eq.\eqn{bh6} and the
$B$ field as
 \be
B_2={m \over 2} \log{r^6-a^6\over r_0^6} ~ \omega_2 \ ,
\ee
where $\omega_2$ is a closed 2-form over 
$S^2$'s
$$\omega_2={1\over2}(\sin\theta_1d\theta_1 \wedge
d\phi_1 -\sin\theta_2d\theta_2 \wedge d\phi_2)  . $$ 
Then the corresponding 3-form field strength becomes
\bea
H_{(3)}= {3 m  r^5\over (r^6-a^6) } dr \wedge \omega_2
\label{bh7}
\eea
where $m$ is a constant parameter. 
The field strength $F_{(3)}$ 
is determined by the Hodge-dual relations \eqn{dual} and it is 
\be
F_{(3)}=m~\omega_3  
\ee
where $\omega_3= (d\psi+\cos\theta_id\phi_i)\wedge \omega_2$.
There is a constant flux through $S^3$
\be
\int_{S^3} F_3= m
\ee
which will be quantized.
Due to the self-dual property \eqn{selfdual}
the field strength $G_3$ is harmonic and the corresponding Euclidean 
energy-momentum tensor identically
vanishes. This has been explicitly checked by us. 
This is the reason that the Einstein equations retain their 
empty-space form $R_{\mu\nu}=0$. 
Also we have
$$F_{\mu\nu\lambda} H^{\mu\nu\lambda}=0$$ 
for this background which is required for axion
 to be vanishing. 
 
However, the field expression
\be\lll{bh10a}
{1\over 2.3!}(H_{\mu\nu\lambda})^2={(9 m)^2\over (r^6-a^6)} 
\ee
is singular at $r=a$.  
Since $(F)^2=(H)^2\sim {1\over r^6}$ for large $r$, the 
asymptotic behavior 
of these $B$-instantons is different from that of the 
regular $4D$ instantons discussed in the last section. 
Note that, the tensor-fields remain 
divergent at $r=a$ even though the metric \eqn{bh6} is regular there. 
The action for the tensor fields diverges when integrated over
the range $a\le r \le\infty$ because of the singularity of the tensor 
fields at $r=a$. So this cannot represent a regular instanton background
unlike Eguchi-Hanson background. 
This singularity at $r=a$ is similar to the 
Klebanov-Strassler conifold singularity where the fields are 
singular at the tip of the cone $r=0$. Whole of this structure reduces to
the singular conifold case when $a=0$. One had to deform the conifold in 
order to avoid the singularity \cite{klebanov}.

\section{Embedding in ten dimensions}
The six-dimensional conifolds can also be embedded in 
ten-dimensional IIB string theory quite elegantly \cite{klebanov}. 
It is of main interest to insert D3-branes on the above resolved conifold 
\eqn{bh6}. 
The full ten-dimensional ansatze are then 
\bea\label{ten1}
&& ds^2=h(r)^{-{1\over2}}(-dt^2+ dx\cdot dx)+h(r)^{1\over2}\left( 
{dr^2\over f(r)}+{r^2\over 9}f(r) (\tilde\psi)^2 
+{r^2\over 6}\sum_{i=1}^2
(d\theta_i^2+ \sin^2\theta_i d\phi_i^2)
 \right) \br
&& F_5= dh^{-1}\wedge dx^0\wedge dx^1\wedge dx^2\wedge dx^3 + 
Q(r) 
\omega_2\wedge \omega_3 \br
&& \Phi=0, ~~~\chi=0\br
&&F_3=m\omega_3, ~~~B_2={m\over 2} \log{r^6-a^6\over r_0^6}\omega_2
\eea
where $f(r)= 1-{a^6\over r^6}$ and we defined $\tilde\psi\equiv d\psi+ \sum_i 
\cos\theta_i d\phi_i$. 
We have to ensure
that  field strength $F_5\equiv dC_4 +B_2\wedge F_3$ satisfies 
its equations of motion
$$dF_5=d\ast_{10} F_5=H_3\wedge F_3 $$
and is also self-dual. 
Thus all field equations  are solved provided 
\bea\label{ph1}
&& M(r)={m^2\over2} \log{r^6-a^6\over r_0^6} \br
&& f r^5 \partial_r h=-Q(r)=-M(r)- c_0   
\eea
where $c_0,~ r_0$ are constants.
An exact solution exists when $a=0$ (unresolved case)
\be\label{ph2}
h(r)= b_0 + {1\over  r^{4}}\left( {c_0\over 4} 
+{3m^2\over4}\log({r\over r_0})+{3m^2\over16} \right)   
\ee
which is  the solution of \cite{klebanov} provided we set 
$c_0=16 \pi g_s N$. In the absence of 3-form flux, $m=0$, 
the 
above solution represents $N$ D3-branes on a singular 
conifold. 

When $a\ne 0$, an exact solution is a bit cumbersome.
 We can however make 
following suitable radial coordinate choice, 
\be\label{bh8} 
\rho^6=r^6-a^6 , ~~~ 
1-{a^6\over r^6}=({\rho\over r})^6,~~~~~
1+{a^6\over \rho^6}=({r\over \rho})^6
\ee
so that the EH-resolved metric \eqn{bh6} in the new coordinates becomes
\be\label{bh9}
ds_{6}^2=
{d\rho^2 +{\rho^2\over9} 
(\tilde\psi)^2\over q(\rho)^{2/3}} + 
{\rho^2\over 6}q(\rho)^{1/3}\sum_{i=1}^2
(d\theta_i^2+ \sin^2\theta_i d\phi_i^2)
\ee
with the function $q(\rho)= 1+a^6/\rho^6$ and the new coordinate range $0\le 
\rho\le\infty$.  The  singular
conifold is obtained  whenever we set  $a=0$.
In these coordinates  10-dimensional metric \eqn{ten1}
will look like
\bea\label{df4}
&& ds^2=h(\rho)^{-1\over2}(dx_\mu^2)+h(\rho)^{1\over2}\left( 
{d\rho^2 +{\rho^2\over9} 
(\tilde\psi)^2\over q(\rho)^{2\over3} }+ 
{\rho^2\over 6}q(\rho)^{1\over3}\sum_{i=1}^2
(d\theta_i^2+ \sin^2\theta_i d\phi_i^2)
\right) 
\eea
 We find that the solution of the
equation \eqn{ph2} in the region $\rho\gg a$ can be expressed as
\be\label{ph3}
h(\rho)= b_0 + {4\pi g_s N\over  \rho^{4}}\left( 1+{4\over 15}{a^6\over 
\rho^6}+
\cdots \right)
+{3m^2\over\rho^4}\left({1\over16}\log({10\rho^4\over \rho_0^4})
+{1\over 150}{a^6\over \rho^6} \log({10\rho^{10}\over \rho_0^{10}})+\cdots 
\right)
\ee
which gives exactly the metric in \cite{klebanov} once $a=0$.
But the conifold is 
the resolved conifold now.
Even in the absence of  
H-flux ($m=0$) the brane geometry does not exactly look like 
 $AdS_5\times T^{1,1}$ if $b_0$ is dropped, that is when $4\pi g_sN\gg \rho$.  
The $AdS$ space will be obtained only when we take 
$4\pi g_sN \gg \rho \gg a$, that is the branes must be located far 
away from the origin of the conifold. Thus if large number of
coincident D3-branes are placed 
in the region $\rho\gg a$, the near horizon geometry 
would become more or less
$AdS_5\times T^{1,1}$. That could be seen by dropping the constant $b_0$
in \eqn{ph3} and setting $m=0$
\be
h(\rho)\sim {4\pi g_s N\over  \rho^{4}}\left( 1+{\cal{O}}({a^6\over 
\rho^6})+
\cdots \right)    
\ee
 When $m \ne 0$ the same will apply 
but instead there will be flux dependent deformations, which are logarithmic,
 of the $AdS_5$ 
geometry with interesting consequences in CFT. 
The D3-Branes on resolved conifolds have several interesting 
results  in the dual CFT \cite{klebanov}.

\section{Summary}

In this short note we  constructed regular six-dimensional conifolds simply
starting from Eguchi-Hanson like ans\"atz and solving the resultant first 
order monopole like equations. We then switched on $F_3$ flux,  the 
field strengths are
found to be singular even for the resolved case. We then construct 
solutions in ten-dimensions and obtain D3-brane solutions which spread 
over EH-resolved conifold. The solutions 
are  singular with the presence of 3-form flux background.
Switching off the flux will make the solutions regular.    
However taking near horizon limit does not immediately give us 
$AdS_5\times T^{1,1}$ space. The $AdS$ geometry exists only for the branes 
located  away from the deformed region of the cone.

While we were finishing this work two simultaneous works have appeared
\cite{klebanov-murugan, chen-cvetic} which also discuss $AdS/CFT$ on the 
resolved $S^2$ and $S^2\times S^2$ -conifolds. 
To add, the generalised Eguchi-Hanson solitons in odd dimensions have 
been constructed in \cite{mann}.

\section*{Acknowledgements}
I wish to thank S. Govindrajan for a fruitful 
discussion on conifold.

\end{document}